\documentclass[twocolumn,floats,aps,showpacs]{revtex4}
\usepackage{times,amsmath,graphics,psfrag,latexsym}

\begin{document}
\title{Depinning of a superfluid vortex line by Kelvin waves}
\author{L.A.K. Donev, L. Hough, and R.J. Zieve}
\address{Physics Department, University of California at Davis}
\begin{abstract}
We measure the interaction of a single superfluid vortex with surface
irregularities. While vortex pinning in superconductors usually becomes weaker
at higher temperatures, we find the opposite behavior.  The pinning steadily
increases throughout our measurement range, from $0.15T_c$ to over $0.5T_c$. 
We also find that moving the other end of the vortex decreases the pinning, so
we propose Kelvin waves along the vortex as a depinning mechanism.
\end{abstract}
\maketitle

Pinning sites can trap vortices in a variety of systems.  In superconductors,
where vortex motion leads to nonzero resistance, a vast amount of work has been
devoted to preventing such motion by introducing defects.  Experimental work
has shown that defects comparable in size to the vortex core make effective pin
centers, and that straight \cite{Civale} or splayed \cite{splayed} line defects
can increase the pin strength.  Yet the mechanisms by which vortices interact
with pin sites remain unresolved.  Experimentally, pinning in superconductors
becomes weaker as temperature increases, in  accord with the general assumption
that depinning occurs through thermal activation over local energy barriers. 
In some materials, a constant rate of vortex motion at low temperature
indicates a crossover to quantum tunneling between pin  sites \cite{HTSQT}.  No
experiments have been able to probe the actual interactions.  Similar issues
appear in studies of neutron stars \cite{Langlois}, where vortex pinning or
friction could account for angular momentum irregularities \cite{Jones}.

Here we report studies of a single vortex line in superfluid helium. We observe
a steady {\em increase} of pinning as temperature increases, for $0.15<T/T_c <
0.6$, and suggest that this unusual temperature dependence comes from the
interaction of the pin site with oscillations along the vortex line. 
Studying one vortex rather than a sizeable collection simplifies some of the
issues in pinning.   Helium is also unusual in that the pinning occurs only at
the end of the vortex, so our vortex interacts with a single pin site.  Surface
pinning dominates in superconductors as well for very clean samples
\cite{clean1, clean2} or those with surface structures such as magnetic
dots \cite{Schuller}.

Our measurements use a brass cylinder, 50 mm long and filled with liquid
helium. A fine superconducting wire, about 12 $\mu$m diameter, is stretched
along the cell, displaced from the cylinder's axis by about 0.8 mm. The data
reported here come from two cells, one drilled with inner diameter 2.9 mm (cell
A) and the other reamed to diameter 3.1 mm (cell B). The walls of cell B appear
smoother.  A pumped ${}^3$He cryostat can cool the cell to 300 mK. We
apply a 250 G field perpendicular to the wire.   A current pulse of a few
milliamperes through the wire excites the wire's vibration by a Lorentz force.
The ensuing motion in the magnetic field induces an emf across the wire, which
we amplify and digitize.

\begin{figure}[b]
\begin{center}
\includegraphics{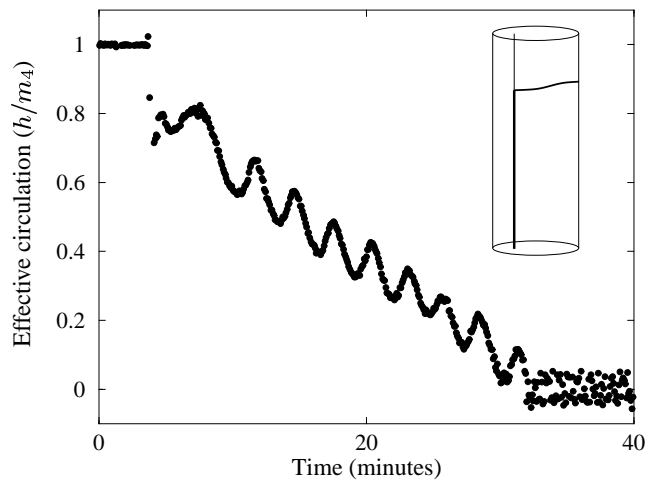}
\caption{\small Unwinding of quantized vortex from wire at 330 mK in cell A.
The inset shows a schematic of the vortex and cell.}
\label{f:1to0}
\end{center}
\end{figure}

In the presence of fluid circulation, vorticity can become trapped on the
wire.  Trapped circulation alters the wire's vibrational normal modes and
splits the fundamental frequency \cite{Vinen,WZ}, causing the plane of
vibration to precess.  A vortex can extend along the entire wire or along only
a portion.  In the latter case, shown in the inset of Figure 1, the vortex core
leaves the wire and traverses the fluid as a free vortex \cite{helicopter,
4Hepaper}.  The splitting of the wire's normal modes due to a partially
attached vortex increases with the length of the vortex on the wire.  We fit
the oscillation envelope to an exponentially damped sine wave to extract the
beat frequency.  We then convert the beat frequency to an effective
circulation, that is, the constant circulation along the wire that would
produce that beat frequency.  For a partial vortex we can convert the beat
frequency to the location of the point where the free vortex meets the
wire.

For the experiments discussed here, we rotate the cryostat to create
circulation but make our measurements, on the behavior of a partially trapped
vortex line, with the cryostat stationary.  The free end of the vortex
precesses around the wire, driven by the flow field of the trapped portion. 
Since the wire is off-center, the point where the free vortex attaches to the
wire moves up and down during the precession to conserve energy.  We detect
this motion from the oscillations it produces in the beat 
frequency \cite{helicopter, 4Hepaper}.  An example is shown in Figure
\ref{f:1to0}.  The oscillation is superimposed on a
steady shortening of the trapped vorticity since the free vortex line
dissipates energy as it moves.  These oscillations occur only at intermediate
circulation, stopping when the vortex is completely on or off the wire.

\begin{figure}[b]
\begin{center}
\includegraphics{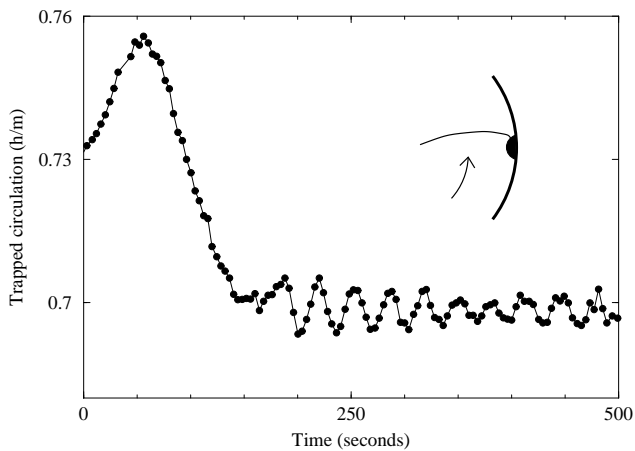}
\caption{\small Kelvin oscillations of a newly pinned vortex at 385 mK.  The
inset illustrates vortex behavior at a bump on the cell wall.}
\label{f:pin}
\end{center}
\end{figure}

As both experiments \cite{Hegde} and computations \cite{Schwarz} have shown,
wall roughness can pin vortices.  A bump on the wall distorts the velocity
field so as to pull the end of the vortex onto the bump.  The vortex may settle
into a metastable pinned state, with its shape determined by the local fluid
velocity, as in the Figure \ref{f:pin} diagram.   A larger velocity field may 
bend the vortex so far to one side that
it moves off the bump entirely.  The critical velocity for working free depends
on the size of the bump \cite{Schwarz, Lukesim}. Figure \ref{f:pin} shows a
pinning event in our cell.  The peak at the left, associated with precession,
suddenly gives way to faster, smaller oscillations around a constant level. The
rapid oscillation has period 33 s and corresponds to the lowest Kelvin
mode, a transverse wave on the free vortex \cite{hell30}.  On an infinite
straight vortex, the Kelvin mode with quarter-wavelength 0.15 cm has period 46
seconds, but the cell geometry alters the period \cite{Lukesim}.  Both the
Kelvin wave along the vortex and the resulting attachment point motion are
clearly visible in our computer simulations of pinning events, and in other
simulations of depinning \cite{Schwarzheli}.

A recent experiment found evidence for single nanometer-scale vortices pinning 
and depinning in a small orifice \cite{AV}.  The critical flow velocity
required to depin the vortex showed no temperature dependence, which was taken
as evidence of quantum tunneling \cite{AV}.  In our experiment, the vortex
frequently precesses without pinning, suggesting that the flow velocity in the
cell is close to the critical velocity for depinning.  Near this critical
velocity, the energy barrier to depinning is small, so one might expect
depinning to be easier at higher temperatures because of thermal
activation.  Our data suggest the opposite: depinning occurs more readily at
{\em low} temperature.

\begin{figure}[tb]
\begin{center}
\includegraphics{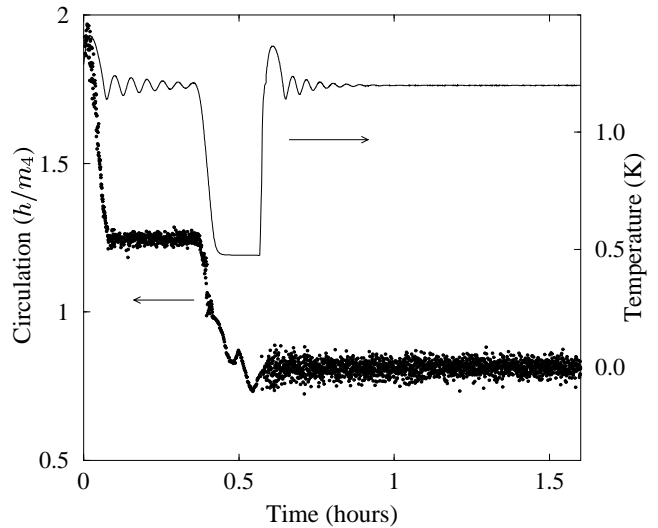}
\caption{\small Circulation (left axis) and temperature (right axis) as a 
function of time.  At
low temperatures the oscillations characteristic of vortex precession
are observed, while at high temperatures neither the oscillations nor
any change in the circulation level can be seen.}
\label{f:freeze}
\end{center}
\end{figure}

Pinning events become more common as temperature increases.  For cell A, we
observed four events in over 40 h of precession measurements below 350 mK, three
events in 8 hours of data between 350 mK and 700 mK, and nine in less than one
hour of vortex motion above 700 mK.  Figure \ref{f:freeze} demonstrates pinning
at our highest temperatures.  The cryostat was rotated immediately before the
data pictured.  The circulation decreases briefly, then levels off at a
nonquantized value and remains there for a full 18 minutes at a temperature
above 1 K, with no sign of vortex precession.  On cooling to 475 mK, the vortex
dislodges and begins its characteristic spiral path through the cell.  Warming
above 1 K again produces a constant level which persists for an hour until the
cryostat warms up.  The increased noise at high temperature hides any Kelvin
oscillations, but the absence of both dissipation and precession-induced
oscillations implies that the vortex is pinned.   In cell A the vortex
consistently pins within minutes of raising the temperature above 700 mK.  In
cell B, with its smoother walls, pinning rarely occurs at any temperature.

\begin{figure}[b]
\begin{center}
\scalebox{.5}{\includegraphics{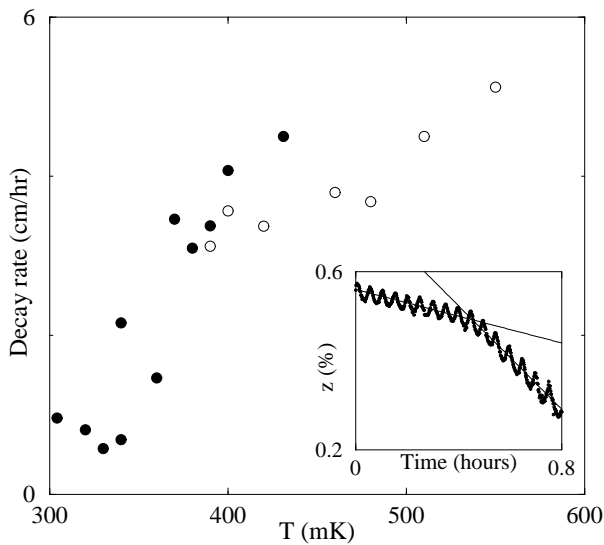}}
\caption{\small Decay rate as a function of temperature (cell B). Filled
circles represent decays from $\bar{\kappa}=1$ to $\bar{\kappa}=0$, while open circles
are from $\bar{\kappa}=2$ to $\bar{\kappa}=1$. Inset: increase in energy dissipation
rate as temperature increases from 340 mK to 380 mK (cell A).  The lines are
guides to the eye.}
\label{f:fricT}
\end{center}
\end{figure}

We next discuss the energy dissipation during vortex precession.  In the
low-temperature regime, where we usually observe precession without pinning, 
the dissipation increases sharply with increasing temperature.  The inset of
Figure \ref{f:fricT} shows a precessing vortex line, with temperature abruptly
changed from 340 mK to 380 mK.  When the temperature goes up, the precession
continues but the background slope increases sharply.  The main portion of
Figure \ref{f:fricT} shows that the decay rate rises steadily with temperature
to above 500 mK.  

The decay rate is significant because the energy dissipation comes from the
same vortex-wall interaction that produces pinning.  One piece of evidence is
that decay rates in cell B are more than a factor of 5 slower than those in
cell A, as expected if cell B has smoother walls.  In fact, the data in the
main part of Figure \ref{f:fricT} are from cell B, since its lower dissipation
reduces the scatter in the higher temperature measurements.   Cell A shows
qualitatively similar behavior.

Interaction of the end of the vortex with the container wall is the only known
mechanism that provides such large dissipation.  Ignoring the displacement
of the wire from the center of the cylinder, the energy stored per length of
the trapped circulation is $u = (\rho_s\kappa^2/4\pi)
\ln(R/R_w)$.  Here $\rho_s$ is the superfluid density,
$\kappa=h/m_4$ is the circulation quantum, $R$ is the cylinder radius,
and $R_w$ is the wire radius.  The vortex of Figure \ref{f:1to0} detaches
completely from the 5 cm long wire in about 30 minutes; so the average
velocity $v_z$ is of order $3\times 10^{-3}$ cm/s and the power dissipated is
$\dot{U}=uv_z\approx 2\times 10^{-10}$ erg/s.

Early experiments on vortex rings measured energy loss due to the scattering of
moving vortices from normal excitations and solvated ${}^3$He atoms \cite{Reif}. 
Typical rings of radius $10^{-4}$ cm, moving at 50 cm/s, lost
``negligible" energy in traveling several cm at 280 mK but up to $2.4\times
10^{-10}$ erg/s at 650 mK.  Our moving vortex has length about 0.15 cm and
velocity about $3\times 10^{-3}$ cm/s.  Since dissipation is proportional to
the vortex length and the square of its velocity \cite{fricreview}, our energy
loss should be more than $10^6$ times {\em smaller} than that of the vortex
rings.  Yet in fact the two are comparable at 650 mK, and at lower 
temperatures our loss is {\em larger}.  Thus normal excitations and
${}^3$He cannot produce our observed damping.

For vortex-boundary interactions, the only previous measurements involve abrupt
changes to a cylindrical container's angular velocity \cite{Adams}.  As the
superfluid returns to equilibrium with the cylinder, vortices move radially,
their ends brushing along the top and bottom walls.  For smooth walls, the
dissipation is consistent with vortices scattering normal fluid entrained by
the cylinder walls. For rough walls, an additional vortex-boundary force,
independent of velocity, is attributed to vortex pinning \cite{Adams}.  The
stored energy per length of a vortex creates a line tension, given by $u_f
\approx (\rho_s\kappa^2/4\pi) \ln(R/a_o)$ for a vortex centered in
a cylinder of radius $R$.  Here $a_o=1.3\times 10^{-8} \mbox{ \AA}$ is the
radius of the free vortex core.  The vortex exerts whatever force is needed to
overcome the pinning, with the restriction that it cannot apply a force greater
than its line tension.  The resulting maximum dissipation is $\dot{U} =
u_f2\pi R/T = 8.8\times 10^{-10}$ erg/s, consistent with our observed
energy loss.

Further evidence for vortex-wall dissipation is that the decay rates are
comparable for $0< \bar{\kappa} <1$ and for $1< \bar{\kappa} <2$. In the latter
case, the circulation about the wire changes from $\kappa=2$  to
$\kappa=1$, with a $\kappa=1$ free vortex undergoing the
precession. Although three times as much energy is lost in going between the
$\kappa=2$ and  $\kappa=1$ states as between $\kappa=1$ and
$\kappa=0$, the larger velocity field increases the precession frequency
by the same factor.  For a velocity-independent force these two effects cancel
so that $v_z$ is unchanged; for a velocity-dependent interaction they do not.

As previously mentioned, the dissipation is stongly temperature dependent. Yet
the only explicit temperature dependence in the line tension mechanism is
through the superfluid density $\rho_s$, which decreases monotonically with
temperature.  At our lowest temperature, $\rho_s$ essentially equals the total
fluid density.  Even at 1 K, $\rho_s$ is within 0.1\% of its low-temperature
limit.  Instead, the observed temperature dependence appears because the
pinning becomes more effective as temperature increases.  As the pinning force
grows stronger, the force countering it from vortex line tension also grows,
and with it the dissipation.  Once the pinning strength exceeds the line
tension, the vortex can no longer pull free, and dissipation vanishes, as shown
in Figure \ref{f:freeze}.  Thus both the temperature dependence of the energy
loss rate and the more direct evidence that vortex motion stops at high
temperature point to an increase in pin strength with temperature.

In cell A the dissipation increases steadily up to 450 mK.  From 450 mK until
the complete pinning begins around 700 mK, the decay rate is too fast for us to
measure accurately.  The highest dissipation observed is about half the line
tension limit, although the circulation disappears so quickly that the error
in identifying the slope is large.  In cell B, where the vortex continues to
move at our highest temperatures, this maximum force is apparently never
needed; and indeed the dissipation never reaches the largest values found in
cell A.

\begin{figure}[tb]
\begin{center}
\psfrag{Decay rate (mum/s)}{\LARGE Decay rate ($\mu$m/s)}
\scalebox{.5}{\includegraphics{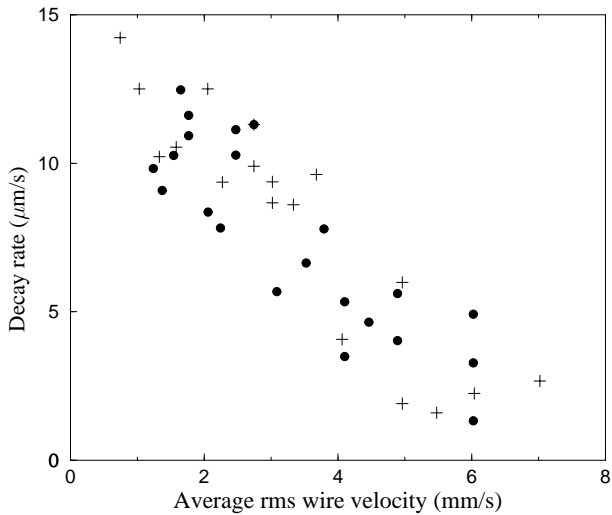}}
\caption{\small Decay rate of trapped vorticity, as a function of average wire 
speed.  +: $T=304-550$ mK, 9 s between pulses.  $\bullet$: $T=400$ mK,
4-20 s between pulses.}
\label{f:fricwire}
\end{center}
\end{figure}

Our explanation is that the wire itself excites vortex waves when we
make our measurements.
As the wire vibrates, typically with an initial amplitude of 10 $\mu$m, the
end of the free vortex lying on the wire is forced to move with it.  This
motion excites a Kelvin wave along the free vortex.  If the oscillation extends
to the far end of the vortex with significant amplitude, it helps the vortex
to wriggle free from pinning sites on the wall.  The damping of the
wire's motion increases sharply with temperature,
making the waves less influential at higher temperatures.  In this 
scenario, the depinning at low temperature seen in Figure \ref{f:freeze}
comes from Kelvin waves shaking the vortex free.

We have verified that motion of the wire reduces dissipation by the moving
vortex.  Figure \ref{f:fricwire} shows the decay rate as a function of the
wire's average rms velocity $v_{av}$.  We calculate $v_{av}$  by integrating
the observed exponential decay,
$$v_{av}=\frac{v\tau(T)}{\Delta t}(1-e^{-\Delta t/\tau(T)}).$$
Here $v$ is the initial rms velocity, $\tau(T)$ the time constant of the wire's
vibration, and $\Delta t$ the interval between measurements. We plot the data
from Figure \ref{f:fricT}, along with data at constant temperature and varying
intervals between pulses. Increasing either temperature or time between wire
pulses corresponds to moving to the left in the figure. The overlay of the two
sets confirms that wire motion is the primary influence on dissipation. 

We suggest that the dissipation during precession comes about as the vortex
encounters wall roughness that distorts it without permanently pinning it. If
the vortex depins upon achieving a critical angle with the wall \cite{Schwarz},
Kelvin waves generated by the wire could help the vortex reach this angle.   As
the vortex leaves a temporary pin site, its  shape is distorted from the usual
precession configuration, giving rise to a line tension component in the
direction of motion.  Uninterrupted, the vortex would oscillate about the wall
normal and average away the work done by vortex line tension. However, if the
vortex encounters another pin site on a time scale shorter than a given mode's 
period, then that mode's contribution will not average away.  The
slowest modes, such as the oscillation of Figure \ref{f:pin} with period 33
s, thus affect the average work from line tension much more
than those generated by the wire, with period near 3 ms. At
high temperatures, the wire and the associated high-frequency modes
contribute less to depinning.  This leads to a larger contribution from
long-wavelength distortion, and more dissipation.

In conclusion, we find an increase in vortex pinning with temperature for a
single vortex pinned only at one end.  We show that the wire's vibration,
mediated by Kelvin waves along the wire, controls vortex depinning. Our further
plans include experiments with smoother and better characterized cell walls,
and tests of the effects of Kelvin oscillations by exciting the wire at
different frequencies between measurements.  Our measurement technique allows a
study of the interaction between a vortex and a single pin site, which
underlies vortex phenomena in superconductors and neutron stars as well as in
liquid helium.

We thank C. Olson, C. Reichhardt, and D. Thouless for helpful conversations,
and UC Davis for funding.

\end{document}